\documentclass[aps,  prd, twocolumn,  superscriptaddress, nofootinbib,
showpacs, amsmath, amssymb]{revtex4}
\usepackage[dvips]{graphicx}

\def\lsim{\lower0.6ex\vbox{\hbox{$ \buildrel{\textstyle <}\over{\sim}\
$}}}      \def\gsim{\lower0.6ex\vbox{\hbox{$      \buildrel{\textstyle
>}\over{\sim}\ $}}}

\def\beq{\begin{equation}} \def\eeq{\end{equation}}

\def\bfig{\begin{figure}[t]   \begin{center}}   \def\efig{\end{center}
\end{figure}}

\def\4he{$^4$He}  \def\7li{$^7$Li}

\def\8Be{$^8$Be}  

\def\Msun{$M_{\odot}$,~}           
   
      \def\rs{r_{\rm s}}  \def\dv2{\Delta_{\rm  V/2}} \def\Rv2{r_{\rm
V/2}}            \def\rhoc{\tilde\rho_{\rm     s}}
\def\Mchi{M_\chi} 

       \def\Msun{\rm       M_{\odot}}
 \def\M9{M_9}

      \def\Mv{M_{\rm   v}}
\def\Vv{V_{\rm   v}}      \def\Rv{R_{\rm   v}}

\def\rs{r_{\rm s}}

\def\photon2{\gamma \gamma} \def\Zgamma{\gamma Z^0} \def\dd{{\rm d}}

\def\LCDM{$\Lambda$CDM}

\def\Deltavir{\Delta_{\rm    vir}}   \def\LCDM{\Lambda    {\rm   CDM}}
\def\cv{c_{\rm v}} \def\GeV{\ {\rm GeV}} \def\cm{{\rm cm}} \def\s{{\rm
s}}    \def\Aeff{{\cal{A}}_{\rm eff}}  \def\LOS{{\rm
LOS}}  \def\TeV{{\rm  TeV}}  \def\texp{t_{\rm  exp}}  \def\ymax{y_{\rm
max}}  \def\Eth{E_{\rm  th}}  \def\Ns{N_{\rm  s}}  \def\NB{N_{\rm  B}}
\def\OmegaMatter{\Omega_{\rm      M}}     \def\OmegaLambda{\Omega_{\rm
\Lambda}} \def\rc{r_{\rm c}} \def\rs{r_{\rm s}} 
\def\RSUSY{\Re_{\rm     Susy}}     \def\LHALO{{\cal{L}_{\rm    halo}}}
   \def\rhoc{\rho_{\rm c}} \def\rhos{\rho_{\rm s}} \def\Mmin{M_{\rm
min}} \def\L{{\cal{L}}}

\begin{document}

\title{The observability of gamma-rays from neutralino annihilations\\
 in Milky Way substructure}

\author{Savvas   M.   Koushiappas}  \email{smkoush@mps.ohio-state.edu}
\author{Andrew    R.    Zentner}%    \email{zentner@cfcp.uchicago.edu}
\altaffiliation[Present  Address: ]{Center  for  Cosmological Physics,
The    University    of    Chicago,    Chicago,   IL    60637,    USA}
\affiliation{Department  of   Physics,  The  Ohio   State  University,
Columbus,    OH    43210,    USA}    \author{Terrence    P.    Walker}
\email{twalker@mps.ohio-state.edu} \affiliation{Department of Physics,
The    Ohio    State   University,    Columbus,    OH   43210,    USA}
\affiliation{Department  of  Astronomy,  The  Ohio  State  University,
Columbus, OH 43210, USA }%

\date{\today}

%%%%%%%%%%%%%%%%%%%%%%%%%%%%%%%%%%%%%%%%%%%%%%%%%%%%%%%%%%%%%%%%%%%%%%%%%%%
\begin{abstract}
We  estimate   the  probability  of  detecting   gamma-rays  from  the
annihilation of  neutralino dark matter  in dense, central  regions of
Milky  Way   substructure.   We  characterize   Galactic  substructure
statistically based on Monte Carlo  realizations of the formation of a
Milky  Way-like  halo  using  a  semi-analytic method  that  has  been
calibrated against the  results of high-resolution N-body simulations.
We find that it may be possible for the upcoming experiments GLAST and
VERITAS,  working in concert,  to detect  gamma-rays from  dark matter
substructure if  the neutralino is relatively light  ($\Mchi \lsim 100
\, \GeV$), while for $\Mchi \gsim  500 \, \GeV$ such a detection would
be unlikely.  We perform most of our calculations within the framework
of  the standard  $\Lambda$CDM  cosmological model;  however, we  also
investigate the  robustness of our results to  various assumptions and
find   that   the   probability   of   detection   is   sensitive   to
poorly-constrained   input   parameters,   particularly   those   that
characterize  the   primordial  power  spectrum.    Specifically,  the
best-fitting power spectrum of the  WMAP team, with a running spectral
index, predicts  roughly a factor  of fifty fewer  detectable subhalos
compared   to   the   standard   $\LCDM$   cosmological   model   with
scale-invariant  power  spectrum.  We  conclude  that  the  lack of  a
detected  gamma-ray signal  gives  very little  information about  the
supersymmetric  parameter space due  to uncertainties  associated with
both the properties of substructure and cosmological parameters.

\end{abstract}
%%%%%%%%%%%%%%%%%%%%%%%%%%%%%%%%%%%%%%%%%%%%%%%%%%%%%%%%%%%%%%%%%%%%%%%%%%%

%\pacs{98.80.-k}

%\vspace{2cm}
\maketitle

%%%%%%%%%%%%%%%%%%%%%%%%%%%%%%%%%%%%%%%%%%%%%%%%%%%%%%%%%%%%%%%%%%%%%%%%%%%
%%%%%%%%%%%%%%%%%%%%%%%%%%%%%%%%%%%%%%%%%%%%%%%%%%%%%%%%%%%%%%%%%%%%%%%%%%%

\section{\label{sec:intro}Introduction}

%%%%%%%%%%%%%%%%%%%%%%%%%%%%%%%%%%%%%%%%%%%%%%%%%%%%%%%%%%%%%%%%%%%%%%%%%%%
%%%%%%%%%%%%%%%%%%%%%%%%%%%%%%%%%%%%%%%%%%%%%%%%%%%%%%%%%%%%%%%%%%%%%%%%%%%

A standard cosmology ($\Lambda$CDM)  has emerged in which the Universe
is  spatially flat and  its energy  budget is  balanced by  $\sim 4$\%
baryonic matter,  $\sim 26$\% cold, collision-less  dark matter (CDM),
and roughly  70\% dark energy  or a cosmological  constant ($\Lambda$)
\cite{wmap,turner,BFPR84}.   The  growth  of  structure is  seeded  by
density  fluctuations supposedly  generated during  an early  epoch of
inflation \cite{BST}.   The primordial power  spectrum of fluctuations
is expected to have a nearly scale-invariant form, $P(k) \propto k^n$,
$n \simeq  1$.  CDM  dominates the baryons  in the matter  budget, and
luminous  galaxies form  within halos  of CDM,  where the  dark matter
potential wells  trap and compress baryons  so that they  may cool and
condense.   In this  paradigm, structure  forms  hierarchically: small
mass objects  collapse first and  merge into larger objects  over time
\cite{WR78}.  Many of the small halos subsumed by objects that grow to
become   present  day  galactic   halos  survive   as  gravitationally
self-bound substructure or ``subhalos.''  In fact, CDM theory predicts
that halos similar to that of  the Milky Way (MW) should host hundreds
of subhalos  that are apparently about  the same size  as the observed
satellite galaxies  of the MW  \cite{K99M99}; however, there  are only
eleven known satellite  galaxies within $\sim 300$ kpc  of the MW (see
Ref.  \cite{MATEO98}   for  a  census  of  the   Local  Group).   This
discrepancy has been dubbed the ``dwarf satellite problem'' (DSP).

The robustness  of the DSP is  uncertain and some  authors have argued
that the observed MW satellites  may correspond directly to the eleven
or  so most  massive subhalos  observed in  simulations \cite{SWTS02}.
Alternatively,  others have proposed  solutions to  the DSP  that fall
into two very broad categories.   In the first, the number of subhalos
of the appropriate size and density structure is reduced such that all
such subhalos  host luminous satellite galaxies.  This  outcome can be
achieved  either  by  changing  the  properties  of  the  dark  matter
\cite{MOD_DM}  or by  modifying the  spectrum of  density fluctuations
that seed  structure growth \cite{KL00,ZB03}.  In the  second class of
solutions, MW-like halos  do play host to hundreds  of subhalos of the
size expected to  host its dwarf satellites, but  luminous galaxies do
not form  in $\sim 90\%$  of these subhalos  and they remain  dark and
undetectable.   In this  case, baryons  do  not cool  and condense  in
subhalos,  perhaps  because  they  are heated  by  the  photo-ionizing
background  \cite{REIONIZE_SQUELCH, PHOTOEVAP}, by  supernova feedback
\cite{SN_FEEDBACK},  or  some other  feedback  mechanism.  Efforts  to
detect  non-luminous substructure  in galactic  halos  through lensing
effects  \cite{STRONG_LENSING}, tidal streams  \cite{TIDAL_STREAM}, or
any  other  method  therefore  promise to  distinguish  between  these
alternatives.  As such, these  observations may provide a crucial test
of the  $\Lambda$CDM paradigm  and reveal important  information about
structure growth and galaxy formation.

If the standard CDM paradigm is correct, then galactic halos should be
teeming with substructure  and if the dark matter is in  the form of a
supersymmetric particle  species produced in the  early Universe, then
it is possible for this substructure to be lit up by the annihilations
of the dark matter particles into gamma-rays.  Supersymmetry (SUSY) is
the  most popular  extension of  the Standard  Model (SM)  of particle
physics (for a review of SUSY, see Ref. \cite{haberkane}) with several
intriguing properties,  one of the most  cosmologically significant of
which is that  it admits natural CDM candidates.   In the most popular
models, especially from the  standpoint of building SUSY grand unified
theories, the  conservation of  R-parity guarantees that  the lightest
supersymmetric particle  (LSP) is stable.  Moreover, there  is a large
swath of SUSY  parameter space in which the  LSP is weakly-interacting
and may  be produced in the  early Universe with  an appropriate relic
density  to  serve as  the  CDM.  In  the  simplest  models that  meet
accelerator constraints, the LSP  is the lightest neutralino ($\chi$),
or the lightest  mass eigenstate formed from the  superposition of the
two CP-even  Higgsinos, the  W$^3$ino, and the  Bino (for a  review of
SUSY  that focuses  on  SUSY dark  matter,  see Ref.  \cite{jungman}).

Several authors have considered the possibility of direct detection of
neutralino dark matter from  the gamma-ray flux produced by neutralino
annihilations in the Galactic center \cite{GALACTIC_CENTER}.  Assuming
that the Milky  Way resides within a standard  NFW-like halo, the flux
from the  Galactic center  may be considerably  larger than  that from
substructure  (by  perhaps  as  much  as  a  factor  of  $\sim  10^3$)
\cite{SWSTY03};  however, there  are large  uncertainties in  the dark
matter  density structure of  Galaxy-sized halos,  largely due  to the
unknown   effect  of   baryons   on  the   dark  matter   distribution
\cite{baryoniceffects}.  In order  to mitigate these uncertainties and
to  test  the  CDM  paradigm,  it  may  be  advantageous  to  consider
annihilations in CDM substructure.  Other authors have considered this
possibility  \cite{bergstrom3,calcaneo,  tasitsiomi,  baltz,  aloisio,
TS03, SWSTY03},  usually with an  eye toward constraining  the minimal
supersymmetric  standard  model  (MSSM) parameter  space.   Generally,
these  studies  have  used  overly optimistic  prescriptions  for  the
spatial distribution and density profiles of CDM subhalos that are not
supported by  detailed studies of structure  formation in $\Lambda$CDM
cosmologies.

In this  paper, we study neutralino annihilation  in halo substructure
using a rather  different approach.  We use a  semi-analytic method to
estimate the  properties of the  subhalo population.  Past  studies of
neutralino  annihilation in  substructure  \cite{bergstrom3, calcaneo,
tasitsiomi,  baltz, aloisio}  did not  account for  the fact  that the
density  structure of  the subhalos  and the  spatial  distribution of
substructure of  a given mass  depend upon their  accretion histories.
The  semi-analytic  model  we  use  allows us  to  account  for  these
correlations, albeit  in an approximate  way.  In addition,  the model
allows us to generate  statistically significant results for a variety
of  input paramters  by examining  a large  number of  realizations of
MW-like  halos.  Thus  we  can estimate  the  likelihood of  observing
gamma-rays from  neutralino annihilations  in dark subhalos.   In this
sense,  our approach  is  aimed more  toward  serving as  a guide  for
observations  and is complementary  to the  recent numerical  study of
Stoehr  et al.  \cite{SWSTY03}.  We  demonstrate that  observations of
this kind  will likely  not yield meaningful  constraints on  SUSY and
explicitly show how  predictions for the expected flux  and the number
of observable subhalos are rather sensitive to uncertainties regarding
the background cosmology.  In  particular, we show that the likelihood
of directly  observing subhalos  via neutralino annihilations  is very
sensitive to  the initial power  spectrum of density  perturbations on
sub-galactic scales.

The   outline  of   this  manuscript   is  as   follows.   In  Section
\ref{sec:DMhalos},  we  review the  properties  of  dark matter  halos
observed  in  N-body simulations  and  discuss  our  modeling of  halo
substructure.   In  Section  \ref{sec:gamma-rays},  we  calculate  the
expected flux from CDM subhalos, describe the various backgrounds that
this signal  must overcome  and outline conditions  for detectability.
In  Section   \ref{sec:results},  we   present  our  results   on  the
detectability   of  substructure   via   gamma-rays  from   neutralino
annihilations.  Lastly,  we summarize  and draw conclusions  from this
work  in Section  \ref{sec:disc}.  Throughout  most of  this  work, we
assume   the   standard   $\Lambda$CDM   cosmological   model   (e.g.,
\cite{wmap,turner}) with $\Omega_{\rm M} = 0.3$, $\Omega_{\rm \Lambda}
= 1 - \Omega_{\rm M} = 0.7$, $\Omega_{\rm B}h^2 = 0.02$, $h=0.72$, and
a  standard,  scale-invariant  primordial  power spectrum  of  initial
density fluctuations, $P(k) \propto k^n$ with $n=1$.  We also consider
the  consequences  of adopting  a  primordial  power  spectrum with  a
running  power  law   index  $\textrm{d}n/\textrm{d}\ln  k  =  -0.03$,
normalized to $\sigma_8 = 0.84$ as advocated by the recent analysis of
the Wilkinson Microwave Anisotropy Probe (WMAP) team \cite{wmap}.

%%%%%%%%%%%%%%%%%%%%%%%%%%%%%%%%%%%%%%%%%%%%%%%%%%%%%%%%%%%%%%%%%%%%%%%%%%%%
%%%%%%%%%%%%%%%%%%%%%%%%%%%%%%%%%%%%%%%%%%%%%%%%%%%%%%%%%%%%%%%%%%%%%%%%%%%%

\section{\label{sec:DMhalos}The structure and substructure of dark matter halos}  

%%%%%%%%%%%%%%%%%%%%%%%%%%%%%%%%%%%%%%%%%%%%%%%%%%%%%%%%%%%%%%%%%%%%%%%%%%%%
%%%%%%%%%%%%%%%%%%%%%%%%%%%%%%%%%%%%%%%%%%%%%%%%%%%%%%%%%%%%%%%%%%%%%%%%%%%%

In this Section, we discuss the density structure of dark matter halos
as  well as the  properties and  characteristics of  halo substructure
based  on  our  semi-analytic  model.   The  size of  a  halo  can  be
quantified  by  its  virial   Mass  $\Mv$,  virial  radius  $\Rv$,  or
equivalently its virial velocity $\Vv^2 \equiv G\Mv/ \Rv$.  The virial
radius  is defined  as the  radius within  which the  mean  density is
$\Deltavir$ times  the mean matter density of  the universe $\rho_{\rm
M}$, so that  $\Mv \equiv 4 \pi \rho_{\rm M}  \Deltavir(z) \Rv^3 / 3$.
The  virial overdensity can  be estimated  using the  approximation of
spherical  top-hat  collapse.   We  compute $\Deltavir(z)$  using  the
approximate  fitting   formula  of  Bryan   and  Norman  \cite{bryan}:
$\Deltavir(z) \simeq  (18 \pi^2 + 82  x - 39  x^2) / \OmegaMatter(z)$,
where  $x  +  1 =  \OmegaMatter(z)=  \OmegaMatter(1+z)^3/[\OmegaMatter
(1+z)^3  + \OmegaLambda]$ and  $\OmegaMatter(z)$ is  the ratio  of the
mean  matter density  to critical  density  at redshift  $z$.  In  the
$\Lambda$CDM cosmology that we  adopt, $\Deltavir(z=0) \simeq 337$ and
$\Deltavir(z)  \rightarrow  178$  at  high redshift,  approaching  the
standard  CDM  ($\OmegaMatter=1$) value.   For  reference, the  virial
radius can be written in terms of  $\Mv$ as $\Rv = 13.6 h^{-1} \, {\rm
kpc}\,  M_9^{1/3}   [\OmegaMatter  \Deltavir(z)  /   337  ]^{-1/3}  \,
(1+z)^{-1}$ where $M_9 = \Mv / 10^9 \, h^{-1} \, \Msun$.

The matter density profiles of CDM halos have been studied extensively
in  numerical simulations.   The spherically-averaged  density profile
proposed by Navarro, Frenk, and White \cite{NFW} (hereafter NFW)
\begin{equation}
\label{eq:NFWprofile}
\rho(r)  =   \rhos  \left(  \frac{r}{\rs}  \right)^{-1}   \left(  1  +
\frac{r}{\rs} \right)^{-2},
\end{equation}
seems  to represent best  the structure  of halos  in the  most recent
studies  of  high-resolution  numerical simulations  \cite{power}  and
seems  to be  further buttressed  by other  theoretical considerations
\cite{dekel_profiles,  TN01}.   These studies  rule  out the  singular
isothermal sphere with $\rho(r)  \propto r^{-2}$ and begin to disfavor
the  very steep profile  with $\rho(r)  \propto r^{-1.5}$  proposed by
Moore et  al. \cite{moore_profile} as faithful  representations of the
predictions   of   CDM  on   small   scales.    In   fact,  Power   et
al.  \cite{power} observe  that  the inner  profiles  of halos  become
increasingly shallow  with decreasing radius  all the way down  to the
minimum  radius   at  which   their  convergence  criteria   are  met.
Therefore, {\it  even the assumption of  the NFW profile  is likely to
overestimate  gamma-ray flux  from dark  matter  annihilation} because
this flux is very sensitive to  the dark matter density at radii below
the  resolution limits  of present  N-body simulations.   A definitive
resolution of  this issue awaits  further numerical work.  In  most of
this  study, we  assume the  NFW profile  with inner  power  law $\rho
\propto  r^{-1}$, but  we  also  explore the  effect  of adopting  the
steeper profile of Moore et al. \cite{moore_profile} on our results.

The  relative  concentration  of  an  NFW  halo of  a  given  mass  is
determined by  the NFW scale  radius $r_{\rm s}$ or  equivalently, the
NFW concentration parameter $\cv  \equiv \Rv/\rs$.  This is related to
the  characteristic NFW  density by  $\rhos =  \rho_{\rm  M} \Deltavir
\cv^3/3 f(\cv)$,  where $f(x) \equiv \ln (1+x)  - x/(1+x)$.  Numerical
studies have  shown that the  concentrations of dark matter  halos are
set  by  their mass  accretion  histories \cite{NFW,ENS,B01,W02}.   In
particular, the concentration  parameter of a halo of  a given mass is
strongly  correlated with  a suitably-defined  epoch of  formation for
typical halos of that mass (see Refs. \cite{NFW,B01,W02} for details).
The  important point  is that  the  structural parameters  of the  NFW
profiles that describe halo  substructure are essentially fixed by the
choice  of cosmology  and are  not set  by, for  example,  requiring a
subhalo  of a  given  mass at  a  given galacto-centric  radius to  be
compact  enough to  resist tidal  stripping  as has  been supposed  in
previous  studies  (e.g., as  in  Ref. \cite{tasitsiomi}).   Moreover,
because  of this  correlation between  subhalo concentration  and halo
formation epoch,  the structural  parameters of halos  depend strongly
upon  cosmological parameters  such as  the linear,  rms  amplitude of
density fluctuations smoothed over  spheres of radius $8$ h$^{-1}$Mpc,
$\sigma_8$, and the power law  index of the primordial power spectrum,
$n$ \cite{D_V2} (of course, if the normalization of the power spectrum
is  fixed on  large scales  by  measurements of  the cosmic  microwave
background  anisotropy,  for  instance,  a  change in  $n$  implies  a
specific  shift in $\sigma_8$).   As we  discuss below,  the gamma-ray
flux from CDM annihilations in  substructure is quite sensitive to the
density  of  the  inner  regions  of halos.   This  implies  that  the
probability of detecting  gamma-ray flux from neutralino annihilations
in substructure  can be a strong function  of cosmological parameters,
even in highly-idealized calculations like those that we present here.

In order to account for  the correlations between halo mass, redshift,
and concentration  we adopt the simple,  semi-analytic prescription of
Bullock et al.  \cite{B01} (B01) in our modeling  of substructure.  We
use this model to set  the concentrations of all substructure halos as
described below.  The B01 model reproduces the mean $\cv-\Mv$ relation
observed  in  N-body  simulations  and  provides an  estimate  of  the
statistical scatter.   The model has been  tested successfully against
standard  $\Lambda$CDM simulations,  tilted  $\Lambda$CDM simulations,
simulations  of so-called  ``standard'' CDM  (i.e., $\Omega_{\rm  M} =
1$), scale-free  power law CDM models \cite{B01,colin},  and warm dark
matter cosmologies  \cite{CAV_WDM} for halos  in the mass  range $10^9
\lsim  \Mv/\textrm{M}_{\odot}  \lsim  10^{14}$.  In  addition,  recent
results from new simulation data  together with a re-analysis of older
simulations suggest that the  B01 model accurately represents the mean
$\cv-\Mv$   relation  all   the  way   down  to   $\Mv  \sim   10^7  \
\textrm{M}_{\odot}$  and also  supports  the universality  of the  NFW
profile \cite{klypinb,COLIN_PC}.

The  first  step  in  our  calculation  is  to  estimate  the  spatial
distribution and structural properties of substructure in the MW halo.
The properties of substructure in  CDM halos are determined through an
endless competition  between subhalo accretion and  destruction due to
tidal  forces and  dynamical friction.   Several authors  have studied
substructure     distributions      in     simulated     CDM     halos
\cite{K99M99,klypina,ghigna,MPA_GROUP},   typically  in  cluster-sized
halos.   In  this  study,  we  adopt  a  complementary,  semi-analytic
approach to estimate the  properties of the substructure population as
described  in Ref.  \cite{ZB03}  (hereafter ZB03).   We  give a  brief
summary of the model below, see Ref. \cite{ZB03} for further detail.

First, we generate  the merger histories of MW-sized  host halos using
the extended Press-Schechter  formalism \cite{EPS}.  In particular, we
employ a  modified version  of the merger  tree algorithm  proposed by
Somerville and Kolatt \cite{SK99}.  This  allows us to generate a list
of masses  and accretion  redshifts for all  halos above a  given mass
threshold $M_{\rm min}$, that merged into the present day MW-like halo
during  its  formation.   Second,   we  assign  each  subhalo  an  NFW
concentration at the accretion event.   To do this, we use the Bullock
et al. \cite{B01}  model to calculate the mean  value of the logarithm
of the concentration parameter,  $\langle \log(\cv) \rangle $, for the
mass of  each subhalo at the  time of accretion.  We  then compute the
actual  value of $\cv$  that we  assign to  each subhalo  by selecting
$\cv$ randomly from a  log-normal distribution with standard deviation
$\sigma(\log(\cv)) = 0.14$.  Bullock et al. \cite{B01} determined this
to be  a good  approximation for the  statistical scatter of  $\cv$ at
fixed  mass in  their simulations.   Finally, we  track  the subhalo's
orbit in  the potential of the  host from the time  of accretion until
today, in a manner similar to that of Taylor and Babul \cite{TB01}, in
order to  determine whether or not  the subhalo is  destroyed by tidal
forces once  incorporated into the  parent halo.  Using this  model we
construct 100 statistical realizations of  a MW-sized halo with $\Mv =
1.4 \times 10^{12} \ \textrm{M}_{\odot}$, using a minimum subhalo mass
of $\Mmin  = 10^5 \  \Msun$, and ten  realizations of a  MW-sized halo
with $\Mmin = 10^4 \ \Msun$.

This technique  yields an estimate  of the substructure  population of
MW-sized  halos  that is  in  approximate  agreement  with the  radial
distributions,   mass    functions,   and   velocity    functions   of
high-resolution  N-body simulations  \cite{ZB03}.  This  method offers
improvement over  some previous estimates  of substructure populations
in  studies  of neutralino  annihilation  in  substructure because  we
account for the known  correlations between mass accretion history and
the structure  of both the  host and its  subhalos and we are  able to
study  a statistically  significant  number of  hosts.  Like  previous
numerical studies \cite{klypina,ghigna,MPA_GROUP}, this method reveals
a core-like behavior in the radial distribution of subhalos themselves
for radii $r \lsim 30-70$ kpc, with the exact value depending upon the
subhalo mass  cut one considers.   This is the well-known  result that
subhalos are anti-biased with respect  to the dark matter due to tidal
effects.  The distribution of  subhalos only follows the $\rho \propto
r^{-3}$ behavior of the mean  dark matter distribution at fairly large
halo-centric radii.  In addition,  our method complements N-body work.
First,  it suffers  from no  inherent resolution  effects such  as the
issue of ``overmerging,'' which may cause simulations to underestimate
the  amount of  substructure near  the centers  of  Galaxy-sized halos
\cite{klypina}.  If  resolution issues are not a  problem in numerical
simulations, then  this method  provides an overestimate  of gamma-ray
flux from subhalos.  This is a conservative approach in the context of
this paper.  Second,  our method allows us to  generate a large number
of MW  halo realizations  so that we  may estimate the  probability of
observing gamma-rays  from neutralino annihilations  in otherwise dark
substructure,  given the scatter  in the  subhalo distribution  in any
given host, and demonstrate  the effects of cosmological parameters on
the probability of detection.

With  the  gross properties  of  the  subhalo  distribution in  place,
another issue that must be dealt with is the distribution of matter in
the  very central  regions  of subhalos.  This  is where  most of  the
luminosity  due to annihilations  originates because  the annihilation
rate scales as the square of the mass density (see below).  There will
be  some limit  to the  central density  of the  halo.   The densities
achieved in  the inner regions of  halos may be limited  simply due to
the fact  that neutralinos in  very dense regions will  annihilate, so
something closer to a nearly  constant density core might be expected.
In  this case,  the core  size is  set by  a competition  between mass
in-fall and neutralino annihilation, so a simple approximation for the
core radius $r_{\rm c,0}$  follows from equating the annihilation rate
to the rate of in-falling material,
\begin{equation}
\label{eq:rcore}
\Mchi /  \rhoc(r_{\rm c,0}) \langle \sigma  | v |  \rangle \sim \sqrt{
\Rv^3 / G \Mv},
\end{equation} 
and solving  for $r_{\rm c,0}$.   In Eq. (\ref{eq:rcore}),  $\Mchi$ is
the mass of  the neutralino and $\langle \sigma | v  | \rangle$ is the
thermally-averaged  total cross  section times  relative  velocity for
neutralino  annihilation.  We assume  that the  density at  all points
interior to $r_{\rm c,0}$ is given  by the NFW profile density at that
point, $\rho_{\rm c} = \rho(r_{\rm  c,0})$.  In actuality, most of the
gamma-ray  luminosity arises from  the inner  core regions,  which are
orders of  magnitude smaller than  has been reliably probed  by N-body
simulations, so there is little justification for assuming that an NFW
profile  holds   at  such  small   radii  \cite{power,SWTS02,SWSTY03}.
Moreover,  there are  many  other phenomena  that could  significantly
affect   the   density   structure    of   halos   on   these   scales
\cite{baryoniceffects}.  Effects  that rely  on the presence  of large
baryonic components in the satellite halos are probably negligible for
the vast majority of subhalos that are otherwise dark.  Other effects,
such as  that of tides  and heating due  to rapid encounters  with the
disk of the host and/or other subhalos are likely to be more important
\cite{REDISTRIBUTE,  TB01}.   In  these   cases,  the  net  result  is
generally  to make  subhalos more  diffuse in  their centers  than our
model predicts,  so we  feel that our  prescription gives  the largest
justifiable  overestimate for  the gamma-ray  luminosity  of subhalos.
Nevertheless,  due to the  uncertainties in  the density  structure of
subhalos at very small radii, in the following Sections we investigate
the effect of  adopting core radii given by $r_{\rm  c} = \beta r_{\rm
c,0}$, where $r_{\rm c,0}$ is the core radius given by the solution to
Eq. (\ref{eq:rcore}).  We allow  $\beta$ to vary between $10^{-2}$ and
$10^8$.

%####################  FIGURE 1  ############################################
%#
%#	SUBHALO PROPERTIES
%#
%############################################################################
\begin{figure}
\resizebox{!}{5.3cm} {\includegraphics{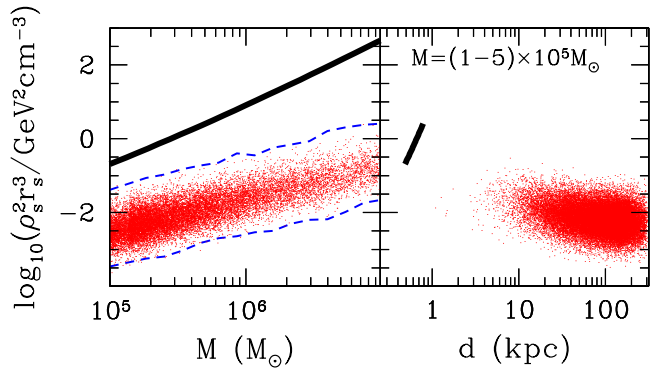}}
\caption{\label{fig:subhalos_model}  Left:  The  relationship  between
$\rho_s^2  \,  r_s^3$  (this   quantity  is  proportional  to  subhalo
luminosity,  see   Section  \ref{sub:subhalo_luminosity}  and  Section
\ref{sec:results}) and the mass  $M$, of Galactic subhalos. The points
represent a scatter  plot of these quantities for  all surviving halos
in one  realization of our  substructure model.  The thick  solid line
represents the  values of  $\rho_s^2 \, r_s^3$  as a function  of mass
using the prescription of Ref. \cite{tasitsiomi} for the properties of
the  subhalos  that they  place  nearest  the  sun. The  dashed  lines
represent  the 99.9  percentile  range of  subhalo  properties in  our
model.   Right: The  relationship  between the  quantity $\rho_s^2  \,
r_s^3$ and  the distance  between the sun  and the subhalo,  $d$.  The
points represent a scatter plot of $\rho_s^2 \, r_s^3$ vs. $d$ for all
of  the  surviving  subhalos  with  masses  in  the  range  $  10^{5}<
M/\textrm{M}_{\odot} < 5 \times 10^{5}$ in ten model realizations. The
thick solid line  represents the range of the  $\rho_s^2 \, r_s^3$-$d$
relationship for subhalos  in the same mass range  obtained by setting
the properties  of the nearest subhalos according  to the prescription
of \cite{tasitsiomi}.  Notice that our modeling suggests that subhalos
have  significantly  lower luminosity  (proportional  to $\rho_s^2  \,
r_s^3$)  and are  typically  further  away than  what  was assumed  in
Ref. \cite{tasitsiomi}. }
\end{figure}
%###########################################################################

Figure   \ref{fig:subhalos_model}  shows  the   relevant  substructure
properties derived  from 100 realizations  of MW-like host  halos.  In
the  left  panel  of  Figure  \ref{fig:subhalos_model},  we  show  the
relationship between  the quantity $\rho_s^2 \, r_s^3$  [as we discuss
below,  this  combination is  roughly  proportional  to the  gamma-ray
luminosity of  the subhalo, see Eq.  (\ref{eq:L_scaling})] and subhalo
mass for the  surviving subhalos in our model.  In  the right panel of
Figure \ref{fig:subhalos_model},  we fix the  mass range to  $ 10^{5}<
M/\textrm{M}_{\odot} <5 \times 10^{5} $ and plot the product $\rho_s^2
\, r_s^3$ as  a function of the distance of the  subhalo from the sun,
$d$.  In both  panels, we also show the values  of $\rho_s^2 \, r_s^3$
that we  computed using the method of  Ref. \cite{tasitsiomi}.  Notice
that our  modeling, which is calibrated against  N-body work, suggests
that  subhalos  are typically  further  away  from  the sun  and  less
luminous  than was  assumed  by Ref.  \cite{tasitsiomi}.  The  subhalo
population that we  present here is a more  faithful representation of
the predictions of  the CDM paradigm of structure  formation.  Yet, we
still expect that all of the simplifying assumptions that we have made
go in the  direction of an over-estimate of  the gamma-ray signal from
neutralino annihilations.

%%%%%%%%%%%%%%%%%%%%%%%%%%%%%%%%%%%%%%%%%%%%%%%%%%%%%%%%%%%%%%%%%%%%%%%%%%%%

%%%%%%%%%%%%%%%%%%%%%%%%%%%%%%%%%%%%%%%%%%%%%%%%%%%%%%%%%%%%%%%%%%%%%%%%%%%%

\section{\label{sec:gamma-rays}The Gamma-Ray signal from annihilations in substructure} 

%%%%%%%%%%%%%%%%%%%%%%%%%%%%%%%%%%%%%%%%%%%%%%%%%%%%%%%%%%%%%%%%%%%%%%%%%%%%
%%%%%%%%%%%%%%%%%%%%%%%%%%%%%%%%%%%%%%%%%%%%%%%%%%%%%%%%%%%%%%%%%%%%%%%%%%%%

In this  Section, we  outline our method  for calculating  the emitted
flux of  gamma-rays from neutralino annihilations in  MW subhalos, and
compare   this  with   the   gamma-ray  background   to  determine   a
detectability threshold.

%%%%%%%%%%%%%%%%%%%%%%%%%%%%%%%%%%%%%%%%%%%%%%%%%%%%%%%%%%%%%%%%%%%%%%%%%%%
\subsection{\label{sub:subhalo_luminosity}Gamma-rays from a subhalo}
%%%%%%%%%%%%%%%%%%%%%%%%%%%%%%%%%%%%%%%%%%%%%%%%%%%%%%%%%%%%%%%%%%%%%%%%%%%%

The  number of  photons with  energy greater  than a  threshold energy
$\Eth$,  arising from  neutralino annihilations  in a  subhalo  can be
written as
\begin{equation}
\label{eq:dNsdEATO}
\Ns =  \frac{1}{4 \pi} \int_{{\Eth}}^{\Mchi}  \frac{\dd \RSUSY(E)}{\dd
E} \, \LHALO(E) \, \Aeff(E) \, \texp \,\dd E,
\end{equation}
where
\begin{equation}
\LHALO(E)  = \int_0^{2  \pi} \dd  \phi  \int_0^{\ymax(E)} y  \, \dd  y
\int_\LOS \frac{\rho^2}{{\cal{D}}^2} \, \dd (\LOS)
\end{equation}
contains all of the information about the subhalo, and
\begin{equation}
\label{eq:susy}
\frac{\dd  \RSUSY}{dE}   =  \sum_{i=\gamma  \gamma,   \gamma  Z^0,  h}
\frac{\dd N_{\gamma,i}}{\dd  E} \frac{\langle  \sigma | v  | \rangle_i
}{\Mchi ^ 2}
\end{equation}
contains   all  the  particle   physics  information.    In  Equations
(\ref{eq:dNsdEATO})-(\ref{eq:susy}),  ${\cal{D}}$ is  the  distance to
the  differential  line-of-sight   ($\LOS$)  element  over  which  the
integral is performed.  The energy-dependent angular resolution of the
detector is $\sigma_{\theta}(E)$ and $d$ is the distance to the center
of the subhalo.  The quantity $\ymax(E)$ represents a projected length
scale on the subhalo defined by the angular resolution of the detector
and given by $\ymax(E)  = \sigma_{\theta}(E) d$.  The energy-dependent
effective  area of  the  detector  is $\Aeff(E)$  and  $\texp$ is  the
exposure time. As discussed above, for $r < r_{\rm c}$, the density is
constant, given by $\rho(r) = \rho(r_{\rm c}) = \rhoc = {\rm const.}$,
while    for   $r   >    r_{\rm   c}$,    $\rho(r)$   is    given   by
Eq. (\ref{eq:NFWprofile}).

The neutralino  mass is $\Mchi$, the  thermally-averaged cross section
times  the relative velocity  of the  annihilating neutralinos  into a
final  state ``$i$'' is  $\langle \sigma  | v  | \rangle_i$,  and $\dd
N_{\gamma,i}/\dd E$  is the number  of photons in the  energy interval
$E$  to $E +  \dd E$  that result  from neutralino  annihilations into
final state ``i.''  The summation runs over the different final states
that lead to gamma-rays.

Neutralino annihilations may  yield photons in three ways:  (1) by the
direct annihilation  into a two-photon final  state ($\gamma \gamma$);
(2) by  the  direct annihilation  into  a  photon  and a  $Z^0$  boson
($\gamma  Z^0$); and  (3)  through annihilation  into an  intermediate
state  that subsequently  decays and/or  hadronizes,  yielding photons
($h$).   The first two  processes are  easy to  deal with  because the
result is a mono-energetic  line emission at energies of $\Eth^{\gamma
  \gamma}=\Mchi$    for    the     $\photon2$    final    state    and
$\Eth^{\Zgamma}=\Mchi  [1 -  (M_{Z^0}/2 \Mchi)^2]$  for  the $\Zgamma$
final state.  In these instances,
\begin{equation}
\frac{\dd  N_{\gamma,k}}{\dd  E}= \frac{n_k}{E}  \,  \delta \left[1  -
\frac{\Eth}{E}\right],
\end{equation}
with   $n_{\photon2}=2$   for   the   $\photon2$  final   state,   and
$n_{\Zgamma}=1$ for the $\Zgamma$  final state.  The theoretical cross
sections for  these processes span  many orders of magnitude  and vary
depending upon  the composition of  the neutralino. In the  context of
the constrained  MSSM, the highest values these  cross sections attain
are  $\sim 10^{-28} \cm^3  \s^{-1}$ when  the neutralino  is primarily
composed  of  higgsinos, while  mostly  gaugino  neutralinos have  the
lowest, roughly  $\sim 10^{-32} \cm^3 \s^{-1}$.  In  order to maximize
the  probability of  detecting gamma-rays  from halo  substructure, we
choose the  most optimistic  set of SUSY  parameters, thus we  fix the
cross  sections for  annihilation  directly into  photons to  $\langle
\sigma |v| \rangle_{\photon2} = \langle \sigma |v| \rangle_{\Zgamma} =
10^{-28} \, \cm^3 \s^{-1}$ for all values of $\Mchi$.

The  much more  complicated case  of  decay and  hadronization of  the
annihilation  products  leads to  continuum  emission.  The  resultant
spectrum  for   the  most  important  annihilation   channels  can  be
well-approximated by \cite{bergstrom1, bergstrom2}
\begin{equation}
\frac{\dd  N_{\gamma,  h}}{\dd   E}  =  \frac{\alpha_1}{\Mchi}  \left(
\frac{E}{\Mchi}    \right)^{-3/2}   {\rm    exp}\left[    -   \alpha_2
\frac{E}{\Mchi} \right],
\end{equation}
where $(\alpha_1 ,\alpha_2) = (0.73,7.76)$ for $WW$ and $Z^0Z^0$ final
states,   $(\alpha_1   ,\alpha_2)   =  (1.0,10.7)$   for   $b\bar{b}$,
$(\alpha_1, \alpha_2)  = (1.1, 15.1)$ for  $t\bar{t}$, and $(\alpha_1,
\alpha_2) = (0.95,6.5)$ for $u\bar{u}$.  The cross sections associated
with these processes also span  many orders of magnitude, from roughly
a few  $\times 10^{-26} {\rm cm}^3 \,  {\rm s}^{-1} $ to  as little as
six orders of magnitude smaller.  Again, we choose the most optimistic
possible value and fix the cross sections to all of these final states
to $\langle \sigma | v | \rangle_h = 5 \times 10^{-26} \ {\rm cm}^3 \,
{\rm s}^{-1}$, independent of the neutralino mass.

%%%%%%%%%%%%%%%%%%%%%%%%%%%%%%
\subsection{Background photons}
%%%%%%%%%%%%%%%%%%%%%%%%%%%%%%

In  order  to  detect  the  signal from  neutralino  annihilations  in
substructure, it  must be significantly larger  than the contaminating
noise  of the gamma-ray  background.  There  are two  contributions to
this background: (1) the observed background due to cosmic ray hadrons
and electrons which will  contribute to the background for atmospheric
{\v{C}}erenkov  telescopes  (ACTs);  and  (2)  the  diffuse  gamma-ray
background  supposedly coming from  astrophysical sources.   The total
contribution to the gamma-ray  count from background photons above the
threshold energy $\Eth$ is
\begin{equation}
\label{eq:background_count}
\NB = \int_{{\Eth}}^\infty \frac{\dd \NB }{\dd E} \, \dd E,
\end{equation}
where  the  number of  observed  background  photons  per unit  energy
interval can be written as
\begin{equation}
\label{eq:backgound_photons}
\frac{\dd  \NB  }{\dd  E}   \simeq  2.67  \,  \Phi_{\rm  B}(E)  \left(
\frac{\Aeff(E)}{\cm^2}   \right)  \left(  \frac{\sigma_\theta(E)}{{\rm
arcmin}}    \right)^2    \left(    \frac{\texp}{{\rm   yr}}    \right)
\,\frac{1}{\textrm{GeV}},
\end{equation}
with
\begin{eqnarray}
\Phi_{\rm  B}(E)  &  =  &  6.4 \times  10^{-2}  \left(  \frac{E}{\GeV}
\right)^{-3.3} +  1.8 \left( \frac{E}{\GeV}  \right)^{-2.75} \nonumber
\\ & + & 1.4 \times 10^{-6} \left( \frac{E}{\GeV} \right)^{-2.20}.
\end{eqnarray}
The three  background contributions to  $\Phi_B$ are as  follows.  The
first  term corresponds to  the electron  background \cite{nishimura},
the second  term is due  to hadronization of cosmic  rays \cite{ryan},
and  the  third  term  accounts  for the  diffuse  gamma-ray  emission
measured  by  the  Energetic  Gamma-Ray Experiment  Telescope  (EGRET)
\cite{sreekumar}.   All three  of these  background  contributions are
relevant for ACTs, while  for space-based gamma-ray detectors the only
relevant  background is  the diffuse  background due  to astrophysical
sources.  A  fourth source  of background photons  can be  the diffuse
emission from neutralino annihilations  in the smooth component of the
MW halo itself;  however, the flux in this case is  more than an order
of magnitude  smaller than all  the aforementioned backgrounds  and we
therefore neglect it in this work.

%%%%%%%%%%%%%%%%%%%%%%%%%%%%%%%%%%%%%%%%%%%%%%%%%%%%%%%%%%%%%%%%%%%%%%%%%%%%%%
\subsection{\label{sec:detectability}The Detectability of Substructure}
%%%%%%%%%%%%%%%%%%%%%%%%%%%%%%%%%%%%%%%%%%%%%%%%%%%%%%%%%%%%%%%%%%%%%%%%%%%%%%

We  now  discuss the  conditions  for  detecting substructure  through
neutralino  annihilations and  the  probability for  observing such  a
signal.   The  background  count  follows Poisson  statistics,  so  it
exhibits  fluctuations of amplitude  $\sim \sqrt{\NB}$.   The quantity
called  the  {\em  significance}  $S  = \Ns  /  \sqrt{\NB}$  therefore
describes  the  likelihood  of  misinterpreting  fluctuations  in  the
background  as the  desired signal  from neutralino  annihilation.  In
order to minimize the  likelihood of counting background fluctuations,
one  must   demand  that  the   significance  be  greater   than  some
predetermined  requirement, $S \ge  \nu$.  We  adopt the  very liberal
detection criterion $\nu = 3$.   Note that both the background and the
source photon  counts are directly proportional to  the exposure time,
so that $S \propto \sqrt{\texp}$.

The next  issues that need to  be addressed are  the optimal threshold
energy  for   observation  and  the  neutralino  mass   at  which  the
probability   for  detection  is   maximized.   These   are  important
considerations  because the  significance is  a function  of threshold
energy  and  neutralino  mass  for  a  fixed  exposure  time,  subhalo
distance,  and subhalo  structure,  $S \equiv  S(> \Eth,\Mchi)$.   For
example, at a fixed neutralino mass, the function $\dd \RSUSY / \dd E$
is generally a decreasing function  of energy while at fixed energy it
is  a  decreasing function  of  the  neutralino  mass.  For  ACTs  and
space-based  gamma-ray  detectors,  $\Aeff(E)$ increases  with  energy
while  $\sigma_\theta(E)$  decreases   with  energy.   The  background
contribution $\Phi_{\rm B}(E)$ is a decreasing function of energy.  We
would like to investigate  the best-case-scenario for detection, so we
choose to  observe at a neutralino  mass and above  a threshold energy
where the  function $S(>  \Eth,\Mchi)$ is maximized.   Generally, $S(>
\Eth,M_{\chi}$) peaks  in the  continuum part of  the spectrum.   As a
result,  it is most  fruitful to  look for  evidence of  the continuum
signal  and then use  follow-up observations  of the  line-emission to
confirm that the signal is, indeed, due to neutralino annihilations.

%########################  FIGURE 2  ########################################
%#
%#	THE SIGNIFICANCE
%#
%############################################################################
\begin{figure}
\resizebox{!}{5.5cm} {\includegraphics{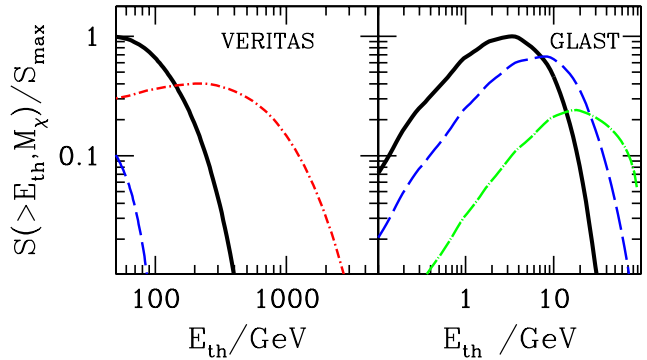}}
\caption{\label{fig:significance} Left:  The significance $S(>  \Eth ,
\Mchi)$ for VERITAS  as a function of threshold  energy for neutralino
masses  $\Mchi=100\GeV$  (long-dashed),  $\Mchi=500\GeV$  (solid)  and
$\Mchi=5\TeV$ (dot-dashed) neutralinos normalized to the peak value of
the $\Mchi=500\GeV$  neutralino case. Right: The  significance for the
GLAST  experiment for  masses $\Mchi=40\GeV$  (solid), $\Mchi=100\GeV$
(dashed)   and  $\Mchi=500\GeV$   (dot-dashed),   normalized  to   the
$\Mchi=40\GeV$ case.   For VERITAS,  the significance is  maximized at
the  instrumental threshold  of $\Eth=50\GeV$  for a  $\Mchi =  500 \,
\GeV$  neutralino,  while for  GLAST  it  is  maximized at  an  energy
threshold of $\Eth=3\GeV$, for a neutralino of mass $\Mchi=40\GeV$.}
\end{figure}
%###########################################################################

For  definiteness, we  consider specific  examples for  maximizing the
detectability of  neutralino annihilations in subhalos by  both an ACT
and  a space-based  gamma-ray detector.   For  the ACT,  we adopt  the
specifications of the Very Energetic Radiation Imaging Telescope Array
System  (VERITAS)  \cite{ong}.   In   this  case  $S$,  and  thus  the
likelihood  of  detection,  is  maximized  for a  $\Mchi  =  500$  GeV
neutralino observed above the VERITAS  energy threshold of $\Eth = 50$
GeV.  We  adopt the specifications  of the Gamma-ray Large  Area Space
Telescope       (GLAST)\footnote{URL       http://glast.gsfc.nasa.gov}
\cite{morselli}  for  the space-based  gamma-ray  detector.  For  this
experiment, $S$ is maximized for a neutralino with mass just above the
present experimental  bounds \cite{PDB},  $\Mchi \approx 40$  GeV, for
gamma-rays  observed above  an energy  threshold  of $\Eth  = 3$  GeV.
Other choices of $\Mchi$ and threshold energy result in smaller values
of the significance.

In Figure  \ref{fig:significance} we show  the detectability condition
normalized to  our choices for  the maximum probability  of detection.
The  flattening  (VERITAS)  and   eventual  drop-off  (GLAST)  in  the
significance at  low energies  is due to  the combined effects  of the
decreasing effective  area of  the detectors as  well as  an increased
background photon  flux. The  lower significance at  higher neutralino
masses results  from the suppression  of flux due to  the $\Mchi^{-2}$
term  in Equation  (\ref{eq:susy}).  In  principle, a  similar  set of
values of the threshold energy and neutralino mass can be obtained for
any detector given its design characteristics.

%%%%%%%%%%%%%%%%%%%%%%%%%%%%%%%%%%%%%%%%%%%%%%%%%%%%%%%%%%%%%%%%%%%%%%%%%%%%%
%%%%%%%%%%%%%%%%%%%%%%%%%%%%%%%%%%%%%%%%%%%%%%%%%%%%%%%%%%%%%%%%%%%%%%%%%%%%%

\section{\label{sec:results}Results}

%%%%%%%%%%%%%%%%%%%%%%%%%%%%%%%%%%%%%%%%%%%%%%%%%%%%%%%%%%%%%%%%%%%%%%%%%%%%%
%%%%%%%%%%%%%%%%%%%%%%%%%%%%%%%%%%%%%%%%%%%%%%%%%%%%%%%%%%%%%%%%%%%%%%%%%%%%%

\begin{figure}
\resizebox{!}{7cm} {\includegraphics{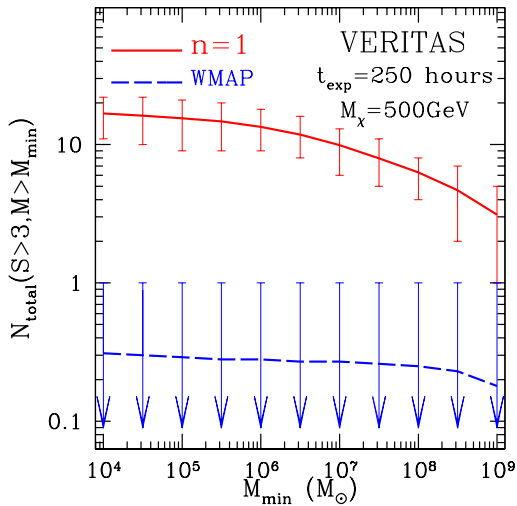}}
\caption{\label{fig:NMmin_VERITAS}  The cumulative number  of subhalos
with mass $M \ge M_{\rm min}$ that are detectable at a significance $S
\ge 3$  by VERITAS on the entire  sky as a function  of $M_{\rm min}$.
As discussed in the text, the  neutralino mass is $\Mchi = 500 \, {\rm
GeV}$ observed  above the  VERITAS threshold energy  of $\Eth =  50 \,
{\rm GeV}$,  yielding the highest possible  detection efficiency.  The
exposure time  is set to  a generous $t_{\rm  exp} = 250$  hours.  The
solid line  represents the mean  over all model realizations,  down to
the  minimum  mass threshold  of  $10^4  \,  \Msun$, in  the  standard
$\Lambda$CDM cosmological model  with scale-invariant primordial power
spectrum.  The   error  bars  demarcate  the   $68$  percentile  range
(symmetric about  the median) determined from  the model realizations.
The dashed line represents the same region for a standard $\Lambda$CDM
model with a ``running'' power law spectrum of primordial fluctuations
with $\textrm{d}n/\textrm{d}\ln  k = -0.03$  as suggested by  the WMAP
team \cite{wmap}.  The down arrows indicate that more than 16\% of the
realizations had  zero detectable  subhalos in the  corresponding mass
bin.}
\end{figure}

Our first results deal with  the likelihood of a detection by VERITAS.
Figure \ref{fig:NMmin_VERITAS} shows the result of our calculation for
the  most optimistic  case of  using VERITAS  to  detect substructure,
namely, a neutralino mass of $\Mchi = 500$ GeV and gamma-rays observed
above the  VERITAS threshold energy  of $\Eth =  50$ GeV.  We  adopt a
detection criterion of  $S \ge 3$ and assume  a very generous exposure
time  of $t_{\rm  exp} =  250$  hours.  In  this Figure,  we show  the
cumulative number of  visible subhalos in the whole  sky as a function
of an adopted lower cut-off in the subhalo mass $\Mmin$.  In practice,
we expect  that there  should be some  cut-off in CDM  substructure at
some small mass,  but any such limit is unknown and  it is expected to
be well beyond the regime  where N-body simulations can probe.  Notice
that the number  of observable subhalos rises only  very slowly as the
minimum mass threshold is reduced.

This  relative flatness  in  the  number of  detectable  objects as  a
function  of  $\Mmin$  can  be  understood from  some  simple  scaling
arguments.   The luminosity  of a  subhalo $\cal{L}$,  can  be roughly
approximated as  the number  of annihilations per  time interval  in a
volume of some  radius $\ymax$ (defined through the  resolution of the
detector)  centered   at  the  center  of  the   subhalo.  Under  this
approximation, there  will be a contribution to  the luminosity coming
from the core and a contribution from the rest of the volume, i.e.,
\begin{equation}
\label{eq:L_scaling}
\L \sim \rhoc^2 \rc^3 +  \int_{\rc}^{\ymax} \rho^2 d^3 r \sim {\cal C}
\, \rhos^2 \rs^3,
\end{equation}
where $ {\cal{C}} = \rc' - \ymax'  $ is a constant that depends on the
core radius through $ \rc'  = \rc / \rs + 1 / ( 1+  \rc / \rs )^3$ and
$\ymax$ through $\ymax'  = 1/(1+ \ymax / r_s)^3$.   Our model yields a
relationship between luminosity and  subhalo mass, $\L \sim M^{0.52}$,
that is close  to what one would expect just  from the simple scalings
$\rhos  \sim \cv^3$,  $\rs \sim  M^{1/3} \,  \cv^{-1}$, and  $\cv \sim
M^{-0.07}$,  which is  the approximate  scaling of  concentration with
mass given by the B01 model  on the small mass scales that we consider
in  this calculation  \cite{B01}.   Knowing how  the  luminosity of  a
subhalo scales  with mass allows  us to derive a  scaling relationship
for the number of visible halos above some flux threshold.  The number
of objects visible above a particular threshold is given by the number
of  objects that  are  contained within  a  volume of  radius $d  \sim
\L^{1/2}$,  such that  the observed  flux is  higher than  the imposed
threshold.  In this  case, the number of objects  above this threshold
per logarithmic mass  interval is $\dd N(\Ns >  \nu \sqrt{N_B})/\dd \,
{\rm ln}M \sim (\dd n / \dd \, {\rm ln} M )\, \L^{3/2}$.  Simulations,
as well as our simple, semi-analytic  model, show that $\dd n / \dd \,
{\rm  ln} M  \sim M^{-0.8}$  \cite{klypina,ghigna}.  Using  the result
that $\L \sim  M^{0.52}$, the number of visible  halos per logarithmic
mass interval is then $\dd N(>  \nu \sqrt{N_B})/ \dd \, {\rm ln}M \sim
M^{-0.02}$ on small  mass scales.  The dependence on  minimum mass cut
is weak: {\em making the mass cut small does not necessarily lead to a
marked increase in the number of visible subhalos.}

%########################  FIGURE 3  ########################################
%#
%#	NMmin_VERITAS
%#
%############################################################################
%###########################################################################

We find that with a lower mass cut at $M_{\rm min} = 10^4 \ \Msun$, we
expect $N_{\rm total}  \sim 17$ subhalos to be detectable  at $S > 3$.
As we discussed  above, considering lower mass cuts  does not increase
the  number   of  detectable   objects  appreciably.   Even   in  this
best-case-scenario, the $68$\% range only extends up to $N_{\rm total}
\simeq  22$  (so  $84$\%  of  the realizations  had  fewer  detectable
subhalos) and the $95 \%$  range extends to $N_{\rm total} \simeq 25$.
This means  that, on  average, a detector  like VERITAS would  have to
survey $\sim 1/20$  of the entire sky to find  {\em one} subhalo; this
amounts  to  one  subhalo  per  $\sim 170$  VERITAS  fields  of  view.
Considering the small field of view  of VERITAS and the fact that ACTs
can,  on  average, observe  only  $\sim 6$  hours  per  night, such  a
detection must rely heavily on serendipitous discovery, even under the
most  favorable of  circumstances.   Note that  for neutralino  masses
higher  or  lower than  $\Mchi  \sim 500$  GeV,  the  total number  of
subhalos detectable at $S > 3$ would be smaller.

If the direct detection of substructure via particle annihilations (or
lack of  such a detection) is  to yield any constraint  on SUSY and/or
information  about  structure/galaxy  formation,  it is  necessary  to
investigate the influence of  the background cosmology on the expected
number   of   visible   subhalos.     As   an   example,   in   Figure
\ref{fig:NMmin_VERITAS} we show the  results of a calculation based on
a cosmology  with $\Omega_{\rm M} =  1 - \Omega_{\rm  \Lambda} = 0.3$,
but  with   a  nonstandard   primordial  power  spectrum   of  density
fluctuations.  In  this particular case, we consider  a power spectrum
with a  running power law index  $dn/d \, {\rm  ln} k = -  0.03$, with
$n(k=0.05  \textrm{ Mpc}^{-1})  =  0.93$, and  $\sigma_8  = 0.84$,  as
suggested by the  recent analysis of the WMAP  data \cite{wmap}.  Note
that  the  statistical significance  of  this  result  is weakened  if
additional uncertainties in the mean flux decrement in the Ly-$\alpha$
forest are considered \cite{SMM03};  nevertheless it is of interest to
investigate  the  consequences of  such  a  power  spectrum.  In  this
alternative model the mean number  of visible subhalos is reduced by a
factor of $\sim  50$.  This effect can be  understood by examining the
influence of  reduced small-scale power  on the average  properties of
CDM   substructure  \cite{NFW,B01,ZB03}.    In  models   with  reduced
small-scale power, low-mass  objects form later, and as  a result have
lower concentrations  than objects  of the same  mass in  the standard
model.  The  reduced concentrations of  typical subhalos and  the fact
that  the   subhalos  are  consequently  more   susceptible  to  tidal
disruption,  and therefore typically  exhibit a  larger core  in their
radial distribution within  the host, results in a  large reduction in
the  number  of  potentially   detectable  subhalos  relative  to  the
predictions of the standard model.

\begin{figure}
\resizebox{!}{7cm} {\includegraphics{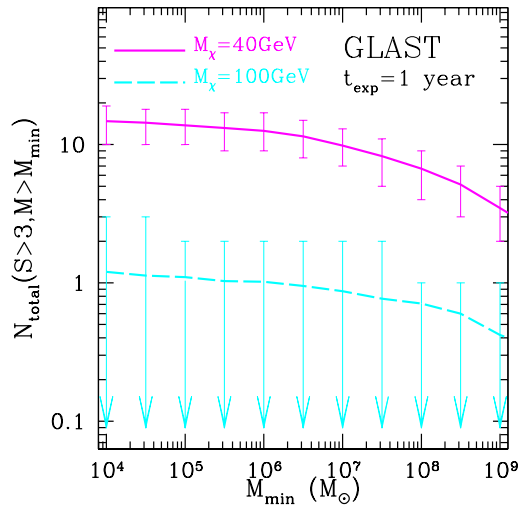}}
\caption{\label{fig:NMmin_GLAST}The  cumulative number of  subhalos on
the entire sky, with mass $M  \ge M_{\rm min}$, that are detectable at
a significance $S \ge 3$ by  GLAST as a function of $M_{\rm min}$.  We
show  results for two  choices of  neutralino mass,  $\Mchi =  40$ GeV
(solid lines) and $\Mchi =  100$ GeV (dashed lines), observed above an
energy  threshold of  $\Eth =  3$ GeV.  The error  bars show  the $68$
percentile range.   As in Figure  \ref{fig:NMmin_VERITAS}, down arrows
indicate  that more  than 16\%  of the  realizations had  zero visible
halos in the corresponding mass bin.}
\end{figure}

We now  turn our attention  to the detectability of  substructure with
GLAST.   In Figure  \ref{fig:NMmin_GLAST},  we exhibit  the number  of
subhalos that  are detectable at  $S > 3$  after a year  long exposure
with GLAST. Here the most  optimistic number of detectable subhalos is
$N_{\rm total}  \sim 19$ at the  upper limit of the  68\% range.  This
amounts to $\sim 2$ objects per field of view, on average; however, it
is necessary to exercise  caution. Consider the energy scales involved
in  this  calculation.   The  optimum  neutralino  mass  for  a  GLAST
detection is at the lower limit of current experimental bounds, $\Mchi
\sim  40$  GeV  \cite{PDB}.   The  factor  $\dd \RSUSY  /  \dd  E$  in
Eq. (\ref{eq:dNsdEATO}) is proportional to $\Mchi^{-2}$ and because of
its  small effective area,  GLAST cannot  compensate for  this rapidly
decreasing  function of  $\Mchi$.  Consequently,  the number  of halos
that are observable drops  dramatically as $\Mchi$ is increased.  This
is    shown   by    the   bottom    set   of    (dashed)    lines   in
Fig. \ref{fig:NMmin_GLAST}.
%#######################   FIGURE 4    #####################################
%#
%#	NMmin_GLAST
%#
%###########################################################################

%###########################################################################

\begin{figure}
\resizebox{!}{7cm} {\includegraphics{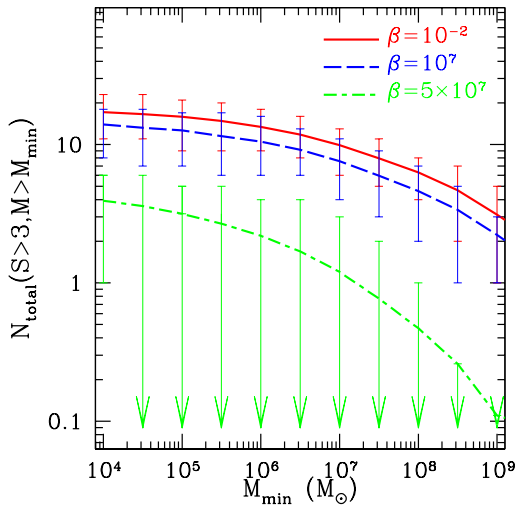}}
\caption{\label{fig:Nbeta}   The  cumulative   number   of  detectable
subhalos with mass $M \ge M_{\rm  min}$ as a function of $M_{\rm min}$
for  different values  of  the quantity  $\beta  = \tilde{r}_{\rm  c}/
\tilde{r}_{\rm c,0}$ (see text).   The neutralino mass, exposure time,
threshold   energy   and   instrument   specifications   are   as   in
Fig. \ref{fig:NMmin_VERITAS}.  Note that  over a change of nine orders
of magnitude in  $\beta$, the total number of  visible halos is nearly
independent of the core size.}
\end{figure}

Due to the  uncertainty in the density structure  of the inner regions
of  halos, it is  of interest  to investigate  the sensitivity  of our
results to the  choice of core radius.  In  Figure \ref{fig:Nbeta}, we
show how  our results on the  number of detectable  subhalos vary with
core radius.  We  parameterize the variation in core  radius by $\beta
\equiv  \rc / r_{\rm  c,0}$, where  $r_{\rm c,0}$  is the  core radius
assigned   according   to    our   standard   procedure   of   solving
Eq. (\ref{eq:rcore}), and  $r_{\rm c}$ is the new  core radius defined
as a multiple of $r_{\rm c,0}$.  Clearly, the particular choice of the
core radius affects our results only  weakly as a shift in core radius
of nine orders of magnitude  changes the number of detectable subhalos
by  less than  40\%.  This  robustness  is easily  understood.  As  we
discussed previously, the luminosity  of a subhalo can be approximated
by   $\L   \sim   \rhos^2   \rs^3    (   \rc'   -   \ymax   ')$   [see
Eq.  (\ref{eq:L_scaling})].   For a  fixed  subhalo  mass  at a  fixed
distance (i.e.,  $\ymax'={\rm const.}$), $\rc'  \approx 1$ as  long as
$\rc  << \rs$  and so  the expected  luminosity  remains approximately
constant as the core radius is  changed.  This behavior is valid up to
approximately $\beta  \sim 10^7$.  At this point,  $\rc ' $  starts to
deviate  from 1,  the  core size  comes  nearer the  typical range  of
$\ymax$  values,  and  the  constant $\cal{C}$  decreases.  Given  the
spatial  distribution of  the  subhalos in  our  model, this  behavior
becomes noticeable at roughly $\rc / r_s \sim 10^{-3} $.  These scales
are  beyond  the current  resolution  limits  of  the inner  parts  of
subhalos  in N-body simulations.   In this  case, our  results suggest
that {\it the presence of smaller cores does not necessarily mean that
the number of potentially detectable subhalos increases appreciably}.

%###########################  FIGURE 5   ####################################
%#
%#	THE EFFECT OF BETA
%#
%############################################################################
%###########################################################################

%%%%%%%%%%%%%%%%%%%%%%%%%%%%%%%%%%%%%%%%%%%%%%%%%%%%%%%%%%%%%%%%%%%%%%%%%%%%
%%%%%%%%%%%%%%%%%%%%%%%%%%%%%%%%%%%%%%%%%%%%%%%%%%%%%%%%%%%%%%%%%%%%%%%%%%%%

\section{\label{sec:disc}Discussion and Conclusions}

%%%%%%%%%%%%%%%%%%%%%%%%%%%%%%%%%%%%%%%%%%%%%%%%%%%%%%%%%%%%%%%%%%%%%%%%%%%%
%%%%%%%%%%%%%%%%%%%%%%%%%%%%%%%%%%%%%%%%%%%%%%%%%%%%%%%%%%%%%%%%%%%%%%%%%%%%

We considered  the detection of  otherwise dark CDM  substructure from
the gamma-ray  products of neutralino  annihilations in the  center of
subhalos.  In order to  demonstrate several important points, we chose
the  most  optimistic  SUSY   parameters  so  that  we  maximized  the
probability  of detection.  We  also employed  a realistic,  yet still
optimistic from  the standpoint of predicting  observable signals from
substructure, model  for the  population of subhalos  in the  MW.  Our
analysis  allowed us to  estimate the  likely range  in the  number of
detectable  subhalos, given  the statistical  fluctuations  in subhalo
populations  from host to  host.  The  simplifying assumptions  of our
model  should only  lead to  overestimates  of the  core densities  of
subhalos  and underestimates of  the typical  distance to  the nearest
subhalo  and,  if  anything,  we  expect that  the  actual  number  of
detectable  subhalos  should  be  smaller  than the  results  that  we
presented here.

%#######################   FIGURE 6    #####################################
%#
%#	Moore profile and n>1 figure
%#
%###########################################################################

\begin{figure}
\resizebox{!}{7cm} {\includegraphics{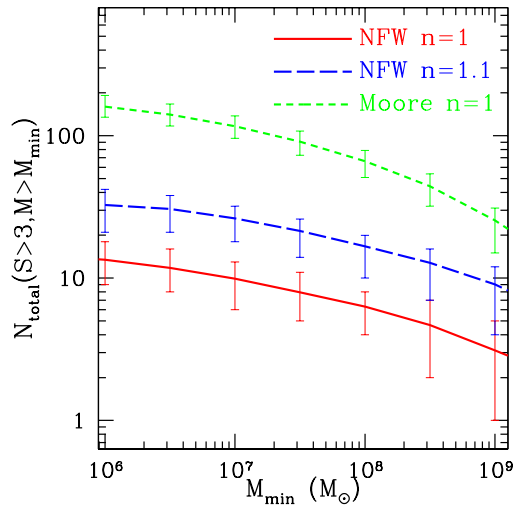}}
\caption{\label{fig:ngt1Moore}The   number   of   subhalos  that   are
detectable by  VERITAS at $S  > 3$ as  a function of the  minimum mass
cut-off $M_{\rm  min}$, after adopting  a Moore profile rather  than a
NFW profile to describe  the substructure (short-dashed) or allowing a
positively  tilted primordial  power  spectrum with  $n=1.1$ and  COBE
normalized  to $\sigma_8  \simeq 1.2$  (long-dashed).  The  solid line
represents the  prediction of our  fiducial ``standard'' model  of NFW
halos  and  a scale-invariant  power  spectrum  with $\sigma_8  \simeq
0.95$.  The error bars are as in Figure \ref{fig:NMmin_VERITAS}.}
\end{figure}
%###########################################################################

We found  it unlikely that an ACT  similar to VERITAS will  be able to
detect  neutralino annihilations  in dark  subhalos due  to  the small
number  of potentially  detectable subhalos  and the  relatively small
field of view of such a detector.  On the other hand, we found that if
the mass  of the neutralino is near  the low end of  the viable range,
$\Mchi \lsim 75$  GeV, the GLAST detector, with  its much larger field
of  view ($\sim  $  2  steradians) will  have  a significantly  higher
probability of detecting such objects.  In our best-case-scenario of a
neutralino  of mass $\Mchi  = 40$  GeV and  an observing  threshold of
$\Eth = 3$ GeV, we found, on average, {\em two} subhalos visible at $S
> 3$  per GLAST field  of view.   Our results  indicate that  the best
strategy  is to  locate gamma-ray  signals with  GLAST and  then point
detectors like  VERITAS at the location  of these signals  in order to
pinpoint them, observe the shape of the gamma-ray spectrum, and search
the energy spectrum  for the line-emission at $E =  \Mchi$, which is a
``smoking  gun'' of  neutralino annihilation  in substructure.   As an
example, we found  that for a neutralino of mass  $\Mchi \sim 75$ GeV,
there will be $\sim 1$ detectable  subhalo per GLAST field of view, on
average,  assuming the  maximum  cross section  for annihilation  into
photons,  and that  a  directed VERITAS  exposure  should confirm  the
line-emission feature after  an exposure time of $\texp  \sim 450$ hr.
If  the  neutralino  mass  were   to  be  in  the  range,  $100  \lsim
\Mchi/\textrm{GeV} \lsim 500$, we found that a detection would require
an instrument with  a large effective area, such  as VERITAS; however,
such  a detection would  have to  rely on  serendipity because  of the
comparably small  field of view of such  detectors. Unfortunately, for
$\Mchi \gsim 500 \, \GeV$ we found that it is unlikely that neutralino
annihilations in dark subhalos will be detectable.

As  we mentioned  earlier, our  results  are rather  sensitive to  the
assumed  form  of the  subhalo  density  profile.   To illustrate  the
significance of  this, we have repeated our  calculations assuming the
density  profile  of Moore  et  al.  \cite{moore_profile}, $\rho(r)  =
\rho_0 (r / r_M)^{-3/2} ( 1 + r / r_M )^{-3/2}$, with $\rho(r) \propto
r^{-3/2}$  at  small  radii.   Our  results are  presented  in  Figure
\ref{fig:ngt1Moore}.   Adopting  the  steeper  Moore  et  al.  profile
\cite{moore_profile}  increases the number  of detectable  subhalos by
approximately  an order  of magnitude.   If subsequent  studies reveal
that  CDM predicts  density  profiles that  vary  as $\rho(r)  \propto
r^{-\eta}$, with  $1 < \eta <  1.5$ intermediate between  the Moore et
al. and  NFW profiles, then  the number of detectable  subhalos should
lie    between     the    solid    and     short-dashed    lines    in
Fig. \ref{fig:ngt1Moore}.

Suppose that gamma-rays are  not observed from halo substructure.  Our
results demonstrate that the lack  of such a detection doesn't lead to
a bonanza of  constraints on SUSY in general,  or the constrained MSSM
in particular, because the number  of subhalos that are detectable via
gamma-rays depends sensitively on  the properties of the substructure.
Even after  choosing the optimal  parameters for detection  (i.e., the
largest couplings to  photons) the likelihood of a  detection is small
for most of the viable mass range of the neutralino.  Additionally, as
we have shown, the likelihood depends upon the density profiles in the
innermost  regions  of  dark  matter  halos.  The  work  of  Power  et
al. \cite{power} shows no  evidence that density profiles approach the
NFW inner power law of  $\rho \propto r^{-1}$.  Rather, they find that
the  power  law may  become  ever  shallower  with decreasing  radius.
Moreover, there  are additional uncertainties that  are not associated
with  our   lack  of  knowledge   of  density  profiles   and  subhalo
populations.  The predictions of  the number of potentially detectable
subhalos are  strongly dependent upon  poorly-constrained cosmological
parameters.  We  showed in Fig.  \ref{fig:NMmin_VERITAS} that adopting
the  best-fitting  power spectrum  from  the  WMAP  group reduces  the
probability of detection by a factor of $\sim 50$ relative to the same
cosmological model  with a standard,  scale-invariant primordial power
spectrum.

Of course, a  detection of gamma-rays from substructure  would yield a
plethora  of information.   First  and foremost,  it  would be  strong
evidence for neutralino  (or some other WIMP which  annihilates into a
photon final  state) dark  matter, it would  indicate the  presence of
numerous,  otherwise dark  subhalos  surrounding the  MW and  disfavor
models with  reduced small-scale power \cite{KL00,ZB03},  and it would
indicate that  such subhalos achieve  high densities in  their central
regions. If  future detections  yield surface brightness  profiles for
substructure, then the dark  matter distribution of the subhalos could
be inferred  which would  be an important  addition to the  studies of
mass  density profiles  \cite{NFW,power,kkbp01,hayashi}.   If a  large
number of  subhalos are found then  that would give  us information on
the  survival of  the  very  inner regions  of  accreted subhalos  and
provide  an  insight  into  the  accretion history  of  the  MW  halo.
However, we must exercise caution:  {\em if} substructure in the Milky
Way halo is  found via annihilation into high-energy  photons, it will
still be  difficult to ``measure''  supersymmetric parameters.  First,
as  we  have already  mentioned,  the  expected  number of  detectable
subhalos    is    sensitive    to   cosmological    parameters.     In
Fig.  \ref{fig:ngt1Moore}, we  show the  number of  subhalos  that are
detectable  by the  VERITAS instrument  in  the case  where the  power
spectrum has a  ``blue tilt,'' with $n=1.1$, and  is normalized to the
Cosmic Background Explorer measurements of cosmic microwave background
anisotropy \cite{cobe}, yielding $\sigma_8  \simeq 1.2$.  In this case
the SUSY parameters  are fixed, yet the number  of detectable halos is
boosted by  a factor of  $\sim 3$ due  to the increase  in small-scale
power.  Although spectra with such  large values of $\sigma_8$ seem to
be disfavored (e.g., \cite{sigma8}),  uncertainties like these must be
marginalized  over  in  analyses  of  the  SUSY  parameter  space  and
calculations similar to the one  presented here may be helpful in this
regard.   Second, the observed  flux of  gamma-rays from  a particular
subhalo depends upon the distance to that particular subhalo.  This is
a  significant problem,  for  there  is no  obvious  way to  determine
reliably the distance to an  otherwise dark subhalo.  In our approach,
we  have   attempted  to   marginalize  over  such   uncertainties  by
calculating ``likely'' realizations of  substructure in the MW and the
most  likely  number  of   observable  subhalos,  assuming  a  maximum
annihilation cross section to photons. Therefore, if a large number of
subhalos are  observed, it is likely that  the neutralino annihilation
cross  section is large.   Such a  large cross  section would  imply a
neutralino  with   a  large  higgsino  component,  a   region  of  the
supersymmetric parameter  space which will be more  easily explored in
forthcoming experiments.  If, on the other hand, SUSY is discovered in
the   upcoming  runs   of  the   Large   Hadron  Collider\footnote{URL
http://lhc-new-homepage.web.cern.ch/lhc-new-homepage/}  (LHC), and the
mass and  couplings of the neutralino are  subsequently measured, this
information  would provide  the framework  within which  a calculation
along the lines of the one  presented in this manuscript could be used
to study the properties and distribution of CDM substructure.

%%%%%%%%%%%%%%%%%%%%%%%%%%%%%%%%%%%%%%%%%%%%%%%%%%%%%%%%%%%%%%%%%%%%%%%%%%%%
%%%%%%%%%%%%%%%%%%%%%%%%%%%%%%%%%%%%%%%%%%%%%%%%%%%%%%%%%%%%%%%%%%%%%%%%%%%%

\begin{acknowledgments}

%%%%%%%%%%%%%%%%%%%%%%%%%%%%%%%%%%%%%%%%%%%%%%%%%%%%%%%%%%%%%%%%%%%%%%%%%%%%
%%%%%%%%%%%%%%%%%%%%%%%%%%%%%%%%%%%%%%%%%%%%%%%%%%%%%%%%%%%%%%%%%%%%%%%%%%%%

We thank Aldo Morselli for providing information on the specifications
of  GLAST.  We thank  James Bullock,  Andrey Kravtsov,  Aldo Morselli,
Stuart  Raby, Leslie Schradin  and Gary  Steigman for  many insightful
conversations. We are grateful to James Bullock and David Weinberg for
helpful comments on an early draft of this manuscript.  SMK thanks the
2003  Santa  Fe  Cosmology  Workshop  where  part  of  this  work  was
completed.  This  research was supported by The  Ohio State University
and by U. S. DOE Contract No. DE-FG02-91ER40690.

\end{acknowledgments}

%%%%%%%%%%%%%%%%%%%%%%%%%%%%%%%%%%%%%%%%%%%%%%%%%%%%%%%%%%%%%%%%%%%%%%%%%%%%
%%%%%%%%%%%%%%%%%%%%%%%%%%%%%%%%%%%%%%%%%%%%%%%%%%%%%%%%%%%%%%%%%%%%%%%%%%%%


\begin{thebibliography}{999}

%%%%%%%%%%%%%%%%%%%%%%%%%%%%%%%%%%%%%%%%%%%%%%%%%%%%%%%%%%%%%%%%%%%%%%%%%%%%
%%%%%%%%%%%%%%%%%%%%%%%%%%%%%%%%%%%%%%%%%%%%%%%%%%%%%%%%%%%%%%%%%%%%%%%%%%%%

\bibitem{wmap}  D. N.  Spergel, L.  Verde, H.  V. Peiris,  E. Komatsu,
M.  R. Nolta,  C.  L. Bennett,  M.  Halpern, G.  Hinshaw, N.  Jarosik,
A. Kogut, M. Limon, S. S. Meyer, L. Page, G. S. Tucker, J. L. Weiland,
E. Wollack and E. L. Wright,  Astrophys. J. Suppl. Ser. {\bf 148}, 175
(2003).

\bibitem{turner}  M. S.  Turner, Int.  J. Mod.  Phys. {\bf  A17}, 3446
(2002).

\bibitem{BFPR84} G.  R. Blumenthal,  S. M. Faber,  J. R.  Primack, and
M. J. Rees, Nature (London) {\bf 311}, 517 (1984).

\bibitem{BST}  J. M.  Bardeen, P.  J.  Steinhardt, and  M. S.  Turner,
Phys. Rev. D, {\bf 28}, 679 (1983).

\bibitem{WR78}    S.    D.    M.     White    and    M.    J.    Rees,
Mon. Not. R. Astron. Soc.  {\bf 183}, 341 (1978).

\bibitem{K99M99}  A. A.  Klypin, A.  V. Kravtsov,  O.  Valenzuela, and
F. Prada,  Astrophys. J.  {\bf 522}, 82  (1999); B. Moore,  S. Ghigna,
F.  Governato,   G.  Lake,  T.   Quinn,  J.  Stadel,  and   P.  Tozzi,
Astrophys. J. Lett. {\bf 524}, L19 (1999).

\bibitem{MATEO98} M. Mateo, Ann. Rev. Astron. and Astrophys. {\bf 36},
435 (1998).

\bibitem{SWTS02}  F.   Stoehr,  S.  D.   M.  White,  G.   Tormen,  and
V. Springel, Mon. Not. R. Astron. Soc. {\bf 335}, L84 (2002).

\bibitem{MOD_DM}  E. D.  Carlson,  M.  E. Machacek,  and  L. J.  Hall,
Astrophys.  J.  {\bf  398},  43  (1992);  D.  N.  Spergel  and  P.  J.
Steinhardt,  Phys. Rev. Lett.  {\bf 84},  3760 (2000);  M. Kaplinghat,
L. Knox,  and M. S.  Turner, Phys. Rev.  Lett. {\bf 85},  3335 (2000);
C. Bento,  O. Bertolami,  R. Rosenfeld, and  L. Teodoro, Phys.  Rev. D
{\bf  62}, 041302  (2000);  T. Matos  and  L. A.  Ure{\~n}a-L{\'o}pez,
Phys. Rev.  D {\b  63}, 063506  (2001); H. Pagels  and J.  R. Primack,
Phys. Rev.  Lett {\bf  48}, 223 (1982);  S. Colombi, S.  Dodelson, and
L.  M. Widrow,  Astrophys. J.  {\bf 458},  1 (1996);  P. Col{\'{\i}}n,
V.  Avila-Reese,  and O.  Valenzuela,  Astrophys.  J.  {\bf 542},  622
(2000); J. J. Dalcanton  and C. J. Hogan, Astrophys.  J. {\bf 561}, 35
(2001); P.  Bode, J.  P. Ostriker,  and N.  Turok, Astrophys.  J. {\bf
556},   93  (2001);  W.   B.  Lin,   D.  H.   Huang,  X.   Zhang,  and
R. Brandenberger, Phys. Rev. Lett. {\bf 86}, 954 (2001).

\bibitem{KL00} M. Kamionkowski and A. R. Liddle, Phys. Rev. Lett. {\bf
84}, 4525 (2000).

\bibitem{ZB03} A.  R.  Zentner and J.  S. Bullock,  Astrophys. J. {\bf
598}, in press (astro-ph/0304292).

\bibitem{REIONIZE_SQUELCH}                G.               Efstathiou,
Mon.  Not. R.  Astron. Soc.  {\bf 256},  43P (1992);  A. A.  Thoul and
D. H.  Weinberg, Astrophys. J. {\bf  465}, 608 (1996);  J. S. Bullock,
A.  V. Kravtsov,  and  D. H.  Weinberg,  Astrophys. J.  {\bf 548},  33
(2001); W.   A.   Chiu,   N.   Y.   Gnedin,  and   J.   P.   Ostriker,
Astrophys.   J.   {\bf   563},    21   (2001);   R.   S.   Somerville,
Astrophys.  J.  {\bf 572},  23  (2002); A.  J.  Benson,  C. S.  Frenk,
C. G. Lacey, C. M. Baugh, and  S. Cole, Mon. Not. R. Astron. Soc. {\bf
333}, 177 (2002).

\bibitem{PHOTOEVAP} R.  Barkana and A. Loeb, Astrophys.  J. {\bf 523},
54 (1999); N. Shaviv and A. Dekel, astro-ph/0305527 (2003).

\bibitem{SN_FEEDBACK} A.  Dekel and J. Silk, Astrophys.  J. {\bf 303},
38  (1986);  G.  Kauffmann,  S.   D.  M.  White,  and  B.  Guiderdoni,
Mon.  Not.  R.   Astron.  Soc.   {\bf  264},  201   (1993);  S.  Cole,
A.  Aragon-Salamaca, C.  S.  Frenk, J.  F.  Navarro, and  S. E.  Zepf,
Mon. Not. R. Astron. Soc. {\bf  271}, 781 (1994); R. S. Somerville and
J. R. Primack, Mon. Not. R. Astron. Soc. {\bf 310}, 1087 (1999).

\bibitem{STRONG_LENSING}    R.    B.    Metcalf    and    P.    Madau,
Astrophys.  J. {\bf  563},  9 (2001);  N.  Dalal and  C. S.  Kochanek,
Astrophys.  J.  {\bf 572},  25  (2001); R.  B.  Metcalf  and H.  Zhao,
Astrophys.   J.  Lett.  {\bf   567},  L5   (2002);  M.   Brada{\v  c},
P. Schneider,  M. Steinmetz, M. Lombardi,  L. J. King,  and R. Porcas,
Astron.  and  Astrophys.  {\bf  388},  373 (2002);  L.  Moustakas  and
R. B. Metcalf, Mon. Not. R. Astron. Soc. {\bf 339}, 607 (2003).

\bibitem{TIDAL_STREAM}   J.   S.   Bullock,   A.  V.   Kravtsov,   and
D. H. Weinberg,  Astrophys. J.  {\bf 548}, 33  (2001); K. V. Johnston,
D.  N. Spergel, and  C. Haydn,  Astrophys. J.  {\bf 570},  656 (2002);
R.   A.  Ibata,   G.   F.  Lewis,   M.   J.  Irwin,   and  T.   Quinn,
Mon.  Not.  R.  Astron. Soc.  {\bf  332},  915  (2002); R.  A.  Ibata,
G.    F.    Lewis,    M.    J.    Irwin,    and    L.    Cambr{\'e}sy,
Mon. Not. R.  Astron. Soc. {\bf 332}, 921 (2002);  L. Mayer, B. Moore,
T. Quinn, F. Governato, and J. Stadel, Mon. Not. R. Astron. Soc.  {\bf
336}, 119 (2002).

\bibitem{haberkane} H. E. Haber and  G. L. Kane, Phys. Rep. {\bf 117},
75 (1985).

\bibitem{jungman}  G.   Jungman,  M.  Kamionkowski,   and  K.  Griest,
Phys. Reports {\bf 267}, 195 (1996).

\bibitem{GALACTIC_CENTER}  V.  Berezinsky,  A.  V.   Gurevich,  and K.
P.  Zybin, Phys.   Lett.   {\bf  B294}, 221  (1992);  P.  Gondolo  and
J. Silk, Phys. Rev. Lett. {\bf  83}, 1719 (1999); G. Bertone, G. Sigl,
and J.  Silk, Mon. Not.  R.   Astron.  Soc.  {\bf 326}, 799 (2001); P.
Ullio, L.  Bergstr{\"o}m,  J.  Edsj{\"o}, and C.  Lacey,  Phys. Rev. D
{\bf 66} 123502 (2002); Hooper \& Dingus, astro-ph/0212509.

\bibitem{bergstrom3} L.  Bergstr{\"o}m, J. Edsj{\"o},  P. Gondolo, and
P. Ullio, Phys. Rev. D, {\bf 59}, 043506 (1999).

\bibitem{calcaneo}   Calc{\'{a}}neo-Rold{\'{a}}n  C.  and   B.  Moore,
Phys. Rev. D, {\bf 62}, 123005 (2000).

\bibitem{tasitsiomi} A. Tasitsiomi and A. V. Olinto, Phys. Rev D, {\bf
66}, 083006 (2002).

\bibitem{baltz}  E. A.  Baltz, C.  Briot, P.  Salati, R.  Taillet, and
J. Silk, Phys. Rev. D {\bf 61}, 023514 (1999).

\bibitem{aloisio}   R.  Aloisio,   P.   Blasi,  and   A.  V.   Olinto,
astro-ph/0206036 (2002).

\bibitem{SWSTY03} F. Stoehr,  S. D. M. White, V.  Springel, G. Tormen,
and N. Yoshida, astro-ph/0307026 (2003).

\bibitem{TS03}      J.      E.      Taylor      and      J.      Silk,
Mon. Not. R. Astron. Soc. {\bf 339}, 505 (2003).

\bibitem{baryoniceffects}  V.  P.   Debattista and  J.   A.  Sellwood,
Astrophys.  J.  {\bf 543}, 704 (2000); J.  A.  Sellwood, Astrophys. J.
{\bf   587},   638   (2003);   O.    Y.  Gnedin,   J.    R.   Primack,
astro-ph/0308385; M.  Milosavljevi{\'{c}},  D.  Merritt, A.  Rest, F.
C.  van den Bosch, Mon.  Not. R.  Astron.  Soc. {\bf 331}, L51 (2002);
D.  Merritt,  M.  Milosavljevi{\'{c}}, L. Verder,  R.  Jimerez, Phys.
Rev.   Lett. {\bf  88}, 191301  (2002); P.   Salucci and  A.  Burkert,
Astrophys.  J.  Lett. {\bf 537}, L9 (2000)

\bibitem{bryan}  G. L.  Bryan and  M. L.  Norman, Astrophys.  J.  {\bf
495}, 80 (1998).

\bibitem{NFW}  J.  F.  Navarro, C.  S.  Frenk,  and  S. D.  M.  White,
Mon.  Not. R.   Astron.  Soc. {\bf  275},  56 (1995);  J. F.  Navarro,
C. S. Frenk, and S. D. M. White, Astrophys.  J. {\bf 462}, 563 (1996);
J. F.  Navarro, C. S. Frenk, and  S. D. M. White,  Astrophys.  J. {\bf
490}, 493 (1997).

\bibitem{power}  C. Power,  J. F.  Navarro, A.  Jenkins, C.  S. Frenk,
S.   D.  M.   White,   V.   Springel,  J.   Stadel,   and  T.   Quinn,
Mon. Not. R. Astron. Soc. {\bf 338}, 14 (2003).

\bibitem{dekel_profiles}  A.   Dekel,  J.  Devor,   and  G.  Hetzroni,
Mon. Not.  R. Astron. Soc. {\bf  341}, 326 (2003); A.  Dekel, I. Arad,
J. Devor, and Y. Birnboim, Astrophys. J. {\bf 588}, 680 (2003).

\bibitem{TN01}  J. E.  Taylor and  J. F.  Navarro, Astrophys.  J. {\bf
563}, 483 (2001).

\bibitem{moore_profile} B.  Moore, T. Quinn, F.  Governato, J. Stadel,
and G. Lake, Mon. Not. R. Astron. Soc. {\bf 310}, 1147 (1999).

\bibitem{ENS}   V.  R.  Eke,   J.  F.   Navarro,  and   M.  Steinmetz,
Astrophys. J. {\bf 554}, 114 (2001).

\bibitem{B01} J. S. Bullock, T. S. Kolatt, Y. Sigad, R. S. Somerville,
A.  V.  Kravtsov,  A.  A.   Klypin,  J.  R.  Primack,  and  A.  Dekel,
Mon. Not. R. Astron. Soc. {\bf 321}, 559 (2001).

\bibitem{W02}  R.   H.  Wechsler,  J.  S.  Bullock,   J.  R.  Primack,
A. V. Kravtsov, and A. Dekel, Astrophys. J. {\bf 568}, 52 (2002).

\bibitem{D_V2}  S. M.  K. Alam,  J. S.  Bullock, and  D.  H. Weinberg,
Astrophys. J. {\bf  572}, 34 (2002); A. R. Zentner  and J. S. Bullock,
Phys. Rev. D {\bf 66}, 043003 (2002); S. S. McGaugh, M. K. Barker, and
W. J. G. de  Blok, Astrophys. J. {\bf 584}, 566 (2003);  F. C. van den
Bosch, H. J. Mo, and X. Yang, astro-ph/0301104 (2003).

\bibitem{colin}  P. Col{\'{\i}}n, A.  A. Klypin,  and A.  V. Kravtsov,
Astrophys. J.  {\bf 539}, 561 (2000).

\bibitem{CAV_WDM} P. Col{\'{\i}}n,  V. Avila-Reese, and O. Valenzuela,
Astrophys. J. {\bf 542},  622 (2000); V. Avila-Reese, P. Col{\'{\i}}n,
O. Valenzuela, E.  D'Onghia, and C. Firmani, Astrophys.  J. {\bf 559},
516 (2001).

\bibitem{klypinb}  A. A.  Klypin, A.  V.  Kravtsov, J.  S. Bullock  \&
J. R. Primack, Astrophys. J. {\bf 554}, 903 (2001).

\bibitem{COLIN_PC} P.  Col{\'{\i}}n, A. Klypin, O. Valenzuela  \& S. ,
S. Gottl{\"o}ber, astro-ph/0308348 (2003).

\bibitem{klypina}  A.  Klypin,  S.  Gottl{\"o}ber,  A.   V.  Kravtsov,
Astrophys. J.  {\bf 516}, 530  (1999); P. Col{\'{\i}}n, A.  A. Klypin,
A.  V.  Kravtsov, and  A. M.  Khokhlov,  Astrophys. J.  {\bf  523}, 32
(1999); J. Chen, A. V. Kravtsov, and C. R.  Keeton, Astrophys. J. {\bf
592}, 24 (2003).


\bibitem{ghigna} S. Ghigna, B. Moore, F. Governato, G. Lake, T. Quinn,
and  J. Stadel,  Mon.  Not. R.  Astron.  Soc. {\bf  300}, 146  (1998);
S. Ghigna  S., B. Moore, F.  Governato, G. Lake, T.  Quinn, J. Stadel,
Astrophys. J. {\bf 544}, 616 (2000).

\bibitem{MPA_GROUP}  F.  Stoehr,  S.  D.  M.  White,  G.  Tormen,  and
V. Springel,  Mon. Not. R. Astron.  Soc. {\bf 335}, L84  (2002); G. De
Lucia, G. Kauffmann, V. Springel, and S. D. M. White, astro-ph/0306205
(2003).

\bibitem{EPS}  J. R.  Bond, S.  Cole,  G. Efstathiou,  and N.  Kaiser,
Astrophys. J. {\bf  379}, 440 (1991); C. Lacey and  S. Cole, Mon. Not.
R. Astron. Soc. {\bf 262}, 627 (1993).

\bibitem{SK99} R. S. Somerville and T. S. Kolatt, Mon. Not. R. Astron.
Soc. {\bf 305}, 1 (1999).

\bibitem{TB01} J. E. Taylor and A. Babul, Astrophys. J. {\bf 559}, 716
(2001); J. E. Taylor and A. Babul, astro-ph/0301612 (2003).

\bibitem{REDISTRIBUTE}   O.    Y.   Gnedin  and    J.   P.   Ostriker,
  Astrophys. J. {\bf  513}, 626 (1999); O. Y.  Gnedin, L. Hernquist and
  J. P. Ostriker, Astrophys. J. {\bf 514}, 109 (1999)

\bibitem{nishimura} J. Nishimura et  al., Astrophys. J. {\bf 238}, 394
(1980).

\bibitem{ryan}  M. J.  Ryan, J.  F. Orme  and V.  K. Balasubrahmanyan,
Phys. Rev. Lett {\bf 28} 985, (1972).

\bibitem{sreekumar} P. Sreekumar et  al., Astrophys. J. {\bf 494}, 523
(1998).

\bibitem{bergstrom1}  L. Bergstr{\"o}m,  P.  Ullio \&  J. H.  Buckley,
Astropart. Phys. 9, 137 (1998)

\bibitem{bergstrom2}   L.  Bergstr{\"o}m,  P.   Salati  \&   J.  Silk,
Nucl. Phys. B346, 129 (1990).

\bibitem{ong} T.  C. Weekes {\it  et al.}, Astropart. Phys.  {\bf 17},
221 (2002).

\bibitem{morselli} A. Morselli, private communication.

\bibitem{SMM03}  U.    Seljak,  P.   V.   McDonald   \&  A.   Makarov,
Mon.  Not. R.  Astron. Soc.  {\bf  342} L79  (2003); R.  A. C.  Croft,
D. H. Weinberg, M. Bolte, S. Burles, L. Hernquist, N. Katz, D. Kirkman
\& D. Tytler, Astrophys.  J. {\bf 581}, 20, (2002).


\bibitem{PDB} D. Groom et al., Phys. Rev. D {\bf 66}, 010001 ({\it The
Review of Particle Physics}) (2002).

\bibitem{kkbp01}  A. A.  Klypin, A.  V. Kravtsov,  J. S.  Bullock, and
J. R. Primack, Astrophys. J. {\bf 554}, 903 (2001).

\bibitem{hayashi} E. Hayashi, J. F.  Navarro, J. E. Taylor, J. Stadel,
and T. Quinn, Astrophys. J. {\bf 584}, 541 (2003).

\bibitem{cobe} C. L. Bennett et al., Astrophys. J. Lett. {\bf 464}, L1
(1994); E. F. Bunn, A. R. Liddle, and M. White, Phys. Rev. D {\bf 54},
5917R (1996);  E. F.  Bunn and  M. White, Astrophys.  J. {\bf  480}, 6
(1997).

\bibitem{sigma8}  M.  Jarvis,  G.  Bernstein, P.  Rischer,  D.  Smith,
B.  Jain, J.  A. Tyson,  and D.  Wittman, Astron.  J. {\bf  125}, 1014
(2003); N. A.  Bahcall et al.,  Astrophys. J.  {\bf 585},  182 (2003);
P.  Schuecker,  H.  B{\"{o}}hringer,  C.  A. Collins,  and  L.  Guzzo,
Astron.  and   Astrophys.  {\bf   398},  867  (2002);   E.  Pierpaoli,
S. Borgani,  D. Scott, and  M. White, Mon.  Not. R. Astron.  Soc. {\bf
342},  163 (2002);  S. Borgani  et al.,  Astrophys. J.  {\bf  561}, 13
(2001).


\end{thebibliography}
\end{document}